# High Temperature Ferromagnetism in $Zn_{1-x}Mn_xO$ semiconductor thin films.

Nikoleta Theodoropoulou*, Vinith Misra, John Philip, Patrick LeClair, Geetha P. Berera and Jagadeesh S. Moodera

*Francis Bitter Magnet Lab, Massachusetts Institute of Technology*

*Cambridge, MA 02139*

Biswarup Satpati and Tapobrata Som

*Institute of Physics, Sachivalaya Marg, Bhubaneswar 751 005, India*

Clear evidence of ferromagnetic behavior at temperatures >400 K as well as spin polarization of the charge carriers have been observed in $Zn_{1-x}Mn_xO$ thin films grown on $Al_2O_3$ and MgO substrates. The magnetic properties depended on the exact Mn concentration and the growth parameters. In well-characterized single-phase films, the magnetic moment is $4.8\mu_B$/Mn at 350 K, the highest moment yet reported for any Mn doped magnetic semiconductor. Anomalous Hall effect shows that the charge carriers (electrons) are spin polarized and participate in the observed ferromagnetic behavior.

In the recent years, significant efforts, both experimental and theoretical, have been focused on the investigation of new magnetic semiconductors.[1,2] Many magnetic semiconductors have been predicted and discovered. The motivation to unravel the mechanism behind ferromagnetism in magnetic semiconductors and to explore the possibility of application of these materials in magneto-electronic devices such as spin-FET (field effect transistors), spin-LED (light emitting diode) or qubits (quantum bits) for quantum computation continues to grow. GaMnAs is the most studied system where the results have been directly compared with theory.[3-5] The highest reported $T_C$ for GaMnAs is 172 K, still far from room temperature.[6] Currently, the main goal is to find a compound semiconductor that is ferromagnetic above room temperature and has a high spin polarization of the charge carriers in order to be used in devices for spin injection and detection. There have been several reports on magnetic semiconductors with a $T_C$ higher than 300 K: GaMnN,[7,8] GaMnP,[9] Co-$TiO_2$,[10-13] $(Cd_{1-x}Mn_x)GeP_2$,[14] $(Zn_{1-x}Mn_x)GeP_2$,[15] and Co-$SnO_2$[16].

The mechanism of ferromagnetism in the III-V based magnetic semiconductors, such as GaMnAs and InMnAs, in relation to their Curie Temperature ($T_C$), magnetic and transport properties is the subject of intensive theoretical investigations.[17-21] The exact $T_C$ depends on the growth recipe, which has yet to be understood and optimized. Most of the theoretical models proposed so far assume Mn ions as the magnetic impurities and consider only p-type carriers, even though not all compound semiconductors can be easily doped (without additional elements) with holes, whereas n-type conductivity is more advantageous in potential technological applications.[22] Recent observations of large anomalous Hall effect in n-type magnetic semiconductors such as Co doped FeSi[23] and $TiO_2$[24] suggest that ferromagnetic behavior is possible in n-type materials and that the existing theoretical models need to be extended. However, the origin of the ferromagnetism in Co doped semiconductors has been questioned since Co clustering cannot be conclusively ruled out. Mn doping is more suitable since Mn and almost all MnOx phases are antiferromagnetic.

ZnO forms in the wurtzite structure, it has a bandgap of 3.32 eV and it is an n-type semiconductor intrinsically because of defects originating from Zn interstitials in the ZnO lattice.[25] $Zn_{1-x}Mn_xO$ can be particularly useful for device applications because of its wide band gap and its transparency to visible light. It can be substitutionally doped with group III elements to create n-type carriers up to $10^{21}$ e/cm$^3$ (desirable for electronic applications) or even be made p-type when doped with group V elements. Dietl et al.[18] predicted that p-type ZnO with 5% Mn should be ferromagnetic at room temperature.

Since the calculation by Dietl et al.,[18] extensive efforts have been focused on producing ferromagnetic ZnO by several growth methods. The results are mixed and diverse: the magnetic behavior seems to be strongly dependent on the particular growth conditions. No ferromagnetism was observed in films grown by pulsed laser deposition (PLD)[26], magnetron sputtering[27] or by solid state reaction method[28], whereas films grown by Laser MBE[29] showed ferromagnetism below 45 K and ZnO single crystals implanted with Mn were ferromagnetic up to 250 K[30]. Recently, while our current work was in progress, weak ferromagnetism above 300 K has been reported for both bulk crystals and thin films of $Zn_{0.98}Mn_{0.02}O$.[31] In this letter we present clear evidence of ferromagnetism in $Zn_{1-x}Mn_xO$ films with a magnetic moment of $5\mu_B$ per $Mn^{+2}$ ion, for x=2.6%, the highest moment yet reported for any magnetic semiconductor doped with Mn. In addition, spin polarization of charge carriers was observed in these thin films with no additional external carrier doping. The estimated Curie temperature is well above room temperature.

The $Zn_{1-x}Mn_xO$ films were grown by reactive magnetron co-sputtering of Zn and Mn targets (dc and rf respectively) in a high vacuum system with a base pressure of about $2\times10^{-6}$ Torr. The substrate temperature was varied from 200 to 380 °C whereas the Oxygen partial pressure was maintained between $1-3\times10^{-4}$ Torr while sputtering with Ar gas at about 20 mTorr. The sputtering power for Zn and Mn were varied to obtain films with different Mn concentrations. A typical deposition rate (obtained by a quartz crystal monitor) was 5 nm/min and the film thickness ranged from 30 to 200 nm. The deposited films were transparent with a slight brown tinge for larger concentrations of Mn (x>0.1). We used several different substrates such as Si(111), quartz, glass, $Al_2O_3$ and MgO(100). However, the magnetic behavior was only observed for films grown on $Al_2O_3$(0001) and MgO(100) substrates and the results presented in this paper are for these films. The crystallinity and phase purity of the films were studied by X-ray diffraction (XRD) and transmission electron microscopy (TEM). The magnetic properties were measured by using a SQUID magnetometer, whereas the optical data were obtained by UV-VIS absorption spectroscopy.

Figure 1 shows the XRD pattern (θ-2θ scan) of a typical $Zn_{1-x}Mn_xO$ film (x=0.05 in this case) on $Al_2O_3$ (0001) substrate. Except for the two characteristic peaks of the ZnO wurtzite structure (002) and (004), all other peaks are due to the sapphire substrate. Thus the films form in single phase with c-axis orientation since the films were deposited on an $Al_2O_3$(0001) substrate.[32] A slight shift of 0.04 (±0.01) degrees of both the (002) and (004) peaks to lower angles was observed in the XRD pattern, compared to pure ZnO (XRD pattern not shown), indicating a c-axis enlargement due to the incorporation of Mn in the ZnO crystal lattice. Such a lattice expansion can be expected since the Mn ionic radius (0.66Å) is larger than that of Zn (0.60 Å).[33] Similar results have been reported on Mn doped ZnO films, although no ferromagnetism was detected.[28] The high resolution cross-sectional TEM image (fig. 2) for a $Zn_{1-x}Mn_xO$ film (x=0.06 in this case) on $Al_2O_3$ (0001) substrate, shows the high crystalline quality of our film as well as no dopant clustering within the TEM resolution, further supporting the single phase growth of the films. The inset of fig.2 shows the selected area diffraction pattern (SADP) of the crystal plane that is oriented perpendicular to the c-axis. The lattice constant *a*, calculated from the SADP is 3.31Å, compared to 3.24 Å for pure ZnO,[25] another indication that Mn is being incorporated in the lattice. It should also be noted that there was no change in the lattice spacing as observed in the SADP taken from different regions confirming that there is no Mn clustering.

The exact Mn concentration, quoted in this report, relatively to Zn was measured by Rutherford Back Scattering (RBS). The films were smooth with a root mean square (rms) surface roughness of less than 2 nm for a 100nm thick film measured by atomic force microscopy. The optical absorption spectrum taken at 295 K showed a clear absorption peak without much of a tail, further showing the integrity of the single-phase films (the presence of even small amounts of additional phases can lead to a broad transition). The bandgap for a 150 nm thick $Zn_{0.96}Mn_{0.04}O$ film was calculated to be 3.28 eV, (from a plot of $\alpha^2E^2$ vs. E, where $\alpha$ is the absorption coefficient and E is the photon energy), in good agreement with the 3.27 eV literature value for ZnO.[25]

The magnetic properties of the films were stable over several months that we tested. The temperature dependence of the field-cooled (FC) magnetization (10 K – 350 K) was taken while warming up, with an applied field of 0.2 T parallel to the plane of a 35 nm thick $Zn_{0.94}Mn_{0.06}O$ film as shown in fig. 3. Similar M(T) curves were observed for all the range of Mn concentrations. The diamagnetic background of the substrate has been subtracted from all data shown. The magnetization of the field cooled sample at 5 K is 2.2 ± 0.4 $\mu_B$/Mn. Although, we do not have any magnetization data above 380 K, the projected $T_C$ is well above room temperature. As a result, the magnetization data in fig. 3 are not true FC data as we cooled down the film starting at 350 K, well below its $T_C$. In order to investigate the magnetic behavior in more detail, one needs to go to higher temperatures. For this reason alone, it is not feasible to draw any further conclusion on the nature of the magnetic interactions, although we note that the M(T) data for a FC sample looks similar to a classical Curie-Weiss curve, at $T<T_C$ as seen in normal ferromagnets.

The magnetization loops taken at 350 K for a 35 nm thick $Zn_{0.95}Mn_{0.05}O$ film is shown in the top inset of fig. 3. The magnetic hysteresis loops for H applied in the film plane or perpendicular to it were similar, showing negligible anisotropy. The coercive field was 50 Oe and hysteresis persisted up to 380 K. The inset at the bottom of fig. 3 shows the saturation magnetic moment, $M_S$, determined by a fit of the M(H) loops at 350K to a Brillouin function and extracting the saturation magnetization at B = 5 T as a function of the Mn concentration for a series of different $Zn_{1-x}Mn_xO$ films. At the lowest Mn concentration, x=2.6%, $M_S$ is 4.8$\mu_B$/Mn at 350 K. For the same film, at low temperatures, $M_S$ is 5$\mu_B$/Mn, according to the M(T) curve shown in fig. 3. This implies that for a g factor of 2, the spin is S = 5/2 which is exactly the value expected for the half-filled d band of divalent Mn ion. The decrease in the $M_S$ with the increase of x is most probably due to the enhancement of the antiferromagnetic interactions between neighboring Mn atoms as the distance between them becomes smaller. Accordingly, the corresponding $T_C$ may be varying with x, a trend that has been observed in the case of GaMnAs where the highest $T_C$ is achieved by x=0.053. It is hard to verify that in our case since the $T_C$ is rather high, exceeding the range of our experimental capabilities. Furthermore, ferromagnetic impurity phases, responsible for the magnetic behavior that we see, can be ruled out since an increase of Mn leads to less, not more magnetism (inset of fig. 3). Additionally, we do not observe any ferromagnetic behavior for bare substrates or pure ZnO and MnOx thin films on $Al_2O_3$ (0001). Since nearly all of the possible MnOx phases that could be forming in our films are antiferromagnetic, with the exception of $Mn_3O_4$, which is ferrimagnetic below 45 K, the observed ferromagnetic behavior above room temperature cannot be due to any second phase formation or impurities in the substrate.

It has been widely believed that in addition to the ferromagnetic behavior, a necessary criterion for the characterization of a dilute magnetic semiconductor is the presence of the Anomalous Hall Effect (AHE)[22,34], as it provides confirmation that the magnetic semiconductor is homogeneous and that the charge carriers are spin polarized. This is observed in our films. Figure 4 shows the Hall resistivity (measured in a Van der Paw configuration) as a function of the magnetic field (H ⊥ to the film plane) for a 50 nm thick $Zn_{0.95}Mn_{0.05}O$ film on MgO(100). The data are taken at 250 K but the same behavior was observed even up to room temperature. The Hall resistance showed a linear dependence at high fields with a negative slope, thus revealing that the conduction mechanism is through electrons (as expected for ZnO). At 250 K the carrier concentration was $4\times10^{19}$ e/cm$^3$, whereas at 300 K, it is $6\times10^{19}$ e/cm$^3$. AHE is believed to originate from a scattering anisotropy of the carriers by embedded local moments due to spin-orbit coupling but its detailed origin is still under investigation. It occurs at low fields, saturating at B = 0.3 T in agreement with the M(H) data (fig. 3) and the coefficient of the AHE is positive, in accordance to what was observed for GaMnAs[4] which is usually explained within a simple two-band model.[35] The electrical resistivity of the films was in the range of 1-500 Ωcm depending on the growth conditions. The inset of fig. 4 shows the temperature dependence of the resistance and carrier concentration showing semiconducting behavior. The carrier concentration (from the slope of Hall resistance at high magnetic fields) slightly varies with temperature, in good agreement with published experimental results for ZnMnO:Al thin films.[37]

The appearance of ferromagnetism was very sensitive to the precise growth conditions such as the substrate temperature (380 °C was the optimum), the exact $O_2$ pressure and the type of substrate; deviations from these conditions led to non-ferromagnetic phases. In particular, the strong ferromagnetic behavior was only seen when the $Zn_{1-x}Mn_xO$ films were grown on sapphire or MgO substrate, whereas films grown on other substrates (quartz, glass, Si) showed weakly paramagnetic behavior. This can explain the varied reports in literature for the magnetic behavior of $Zn_{1-x}Mn_xO$ films.

In the recently reported magnetization data for $Zn_{1-x}Mn_xO$ thin films and bulk,[31] the observed moment is much less: $0.16\mu_B$ /Mn as compared to the $5\mu_B$ /Mn that we observe. The fused quartz substrate used in their work may provide an explanation. Additionally, in that work,[31] no shift on the XRD data was reported for $Zn_{1-}$ $_x$Mn$_x$O as compared to pure ZnO, whereas, they showed from electron energy-loss spectroscopy (EELS) that Mn is in the Mn$^{2+}$ state. According to that result and with g=2, one would expect to see 5$\mu_B$/Mn for optimized conditions, exactly what we see, clearly indicating the Mn ion is in the 2+ state. An uneven distribution of Mn ions in the lattice leading in AF interactions, thereby reducing the total magnetic moment may be one of the reasons for the reduced moment that we see for higher Mn concentrations. One can envision the independent control of magnetization and carrier concentration by tuning the Mn concentration to achieve the desired magnetic behavior and by co-doping ZnO with elements such as Al or Ga to change the carrier concentration. Aditionally, MgO(100) can be epitaxially grown on Si; over which ferromagnetic ZnO:Mn can be grown for Si based electronics.

In conclusion, we provide strong evidence that ZnO doped with Mn is ferromagnetic at temperatures significantly above room temperature and that the carriers are spin polarized electrons that participate in the magnetism. The magnetic moment deduced from our data is the highest seen so far, same as the expected value of 5$\mu_B$/Mn ion. Zn$_{1-x}$Mn$_x$O appears to be a promising candidate as a magnetic semiconductor for use in spin based electronic applications of the future.

\* e-mail: nineta@mit.edu

**Acknowledgments**

We would like to thank Dr. T. H. Kim and T.S. Santos for valuable discussions and experimental assistance. We gratefully acknowledge Dr. P.V. Satyam for extending the TEM facility at IOP Bhubaneswar (India) for this work. This work was supported by a grant from the Cambridge-MIT joint research project and partially by NSF grant DMR-0137632 and NSF-NIRT grant 2003-1266.

**Figure Captions**

Fig. 1: XRD pattern of an $Al_2O_3$ (0001) substrate (top figure) and a $Zn_{0.95}Mn_{0.05}O$ film on $Al_2O_3$ (0001) substrate (bottom figure) showing two ZnO peaks, (002) and (004). No additional peaks due to formation of secondary phases can be observed.

Fig. 2: High-resolution cross-sectional TEM image for a film of $Zn_{0.94}Mn_{0.06}O$. Within the TEM resolution no Mn clustering is observed. The inset shows the SADP of the crystal plane that is oriented perpendicular to the c-axis.

Fig. 3: Field cooled magnetization variation with temperature for a 100nm thick film of $Zn_{0.96}Mn_{0.04}O$; data taken with H= 0.2 T applied in the film plane. The top inset shows the hysteresis loop taken at 350 K for a 35 nm thick $Zn_{.95}Mn_{.05}O$ film. The bottom inset shows the dependence of the saturation magnetization on the Mn concentration at T=350 K.

Fig. 4: The Hall resistance of a $Zn_{0.96}Mn_{0.04}O$ film grown on MgO (100) at 250 K. B is perpendicular to the film plane. The anomalous Hall effect is clear at low fields. At high fields the normal Hall effect dominates. Inset: Temperature dependence of the resistivity and the carrier concentration of the film.

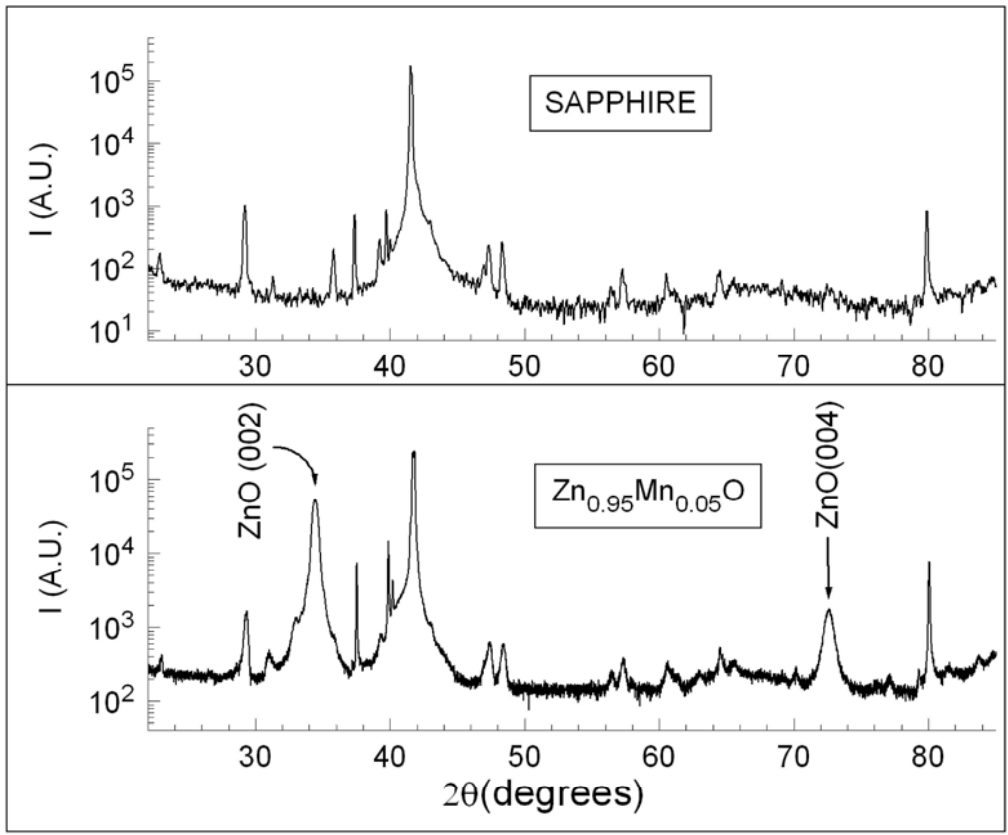

Figure 1

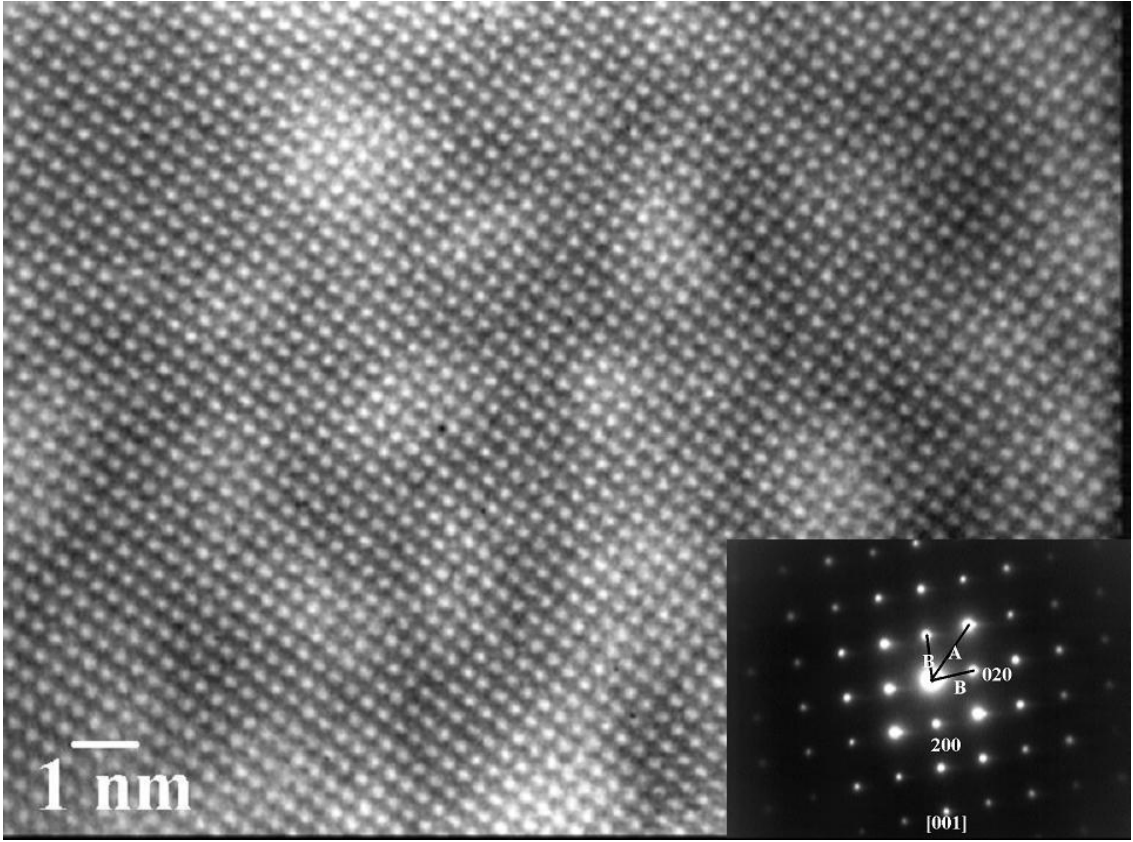

Figure 2

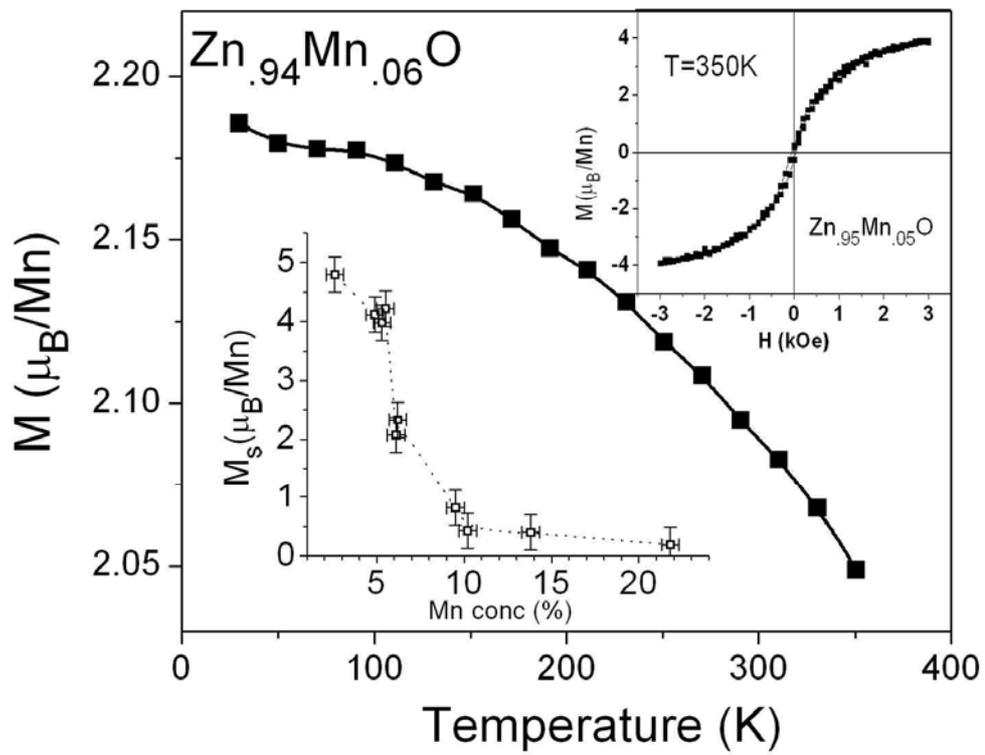

Figure 3

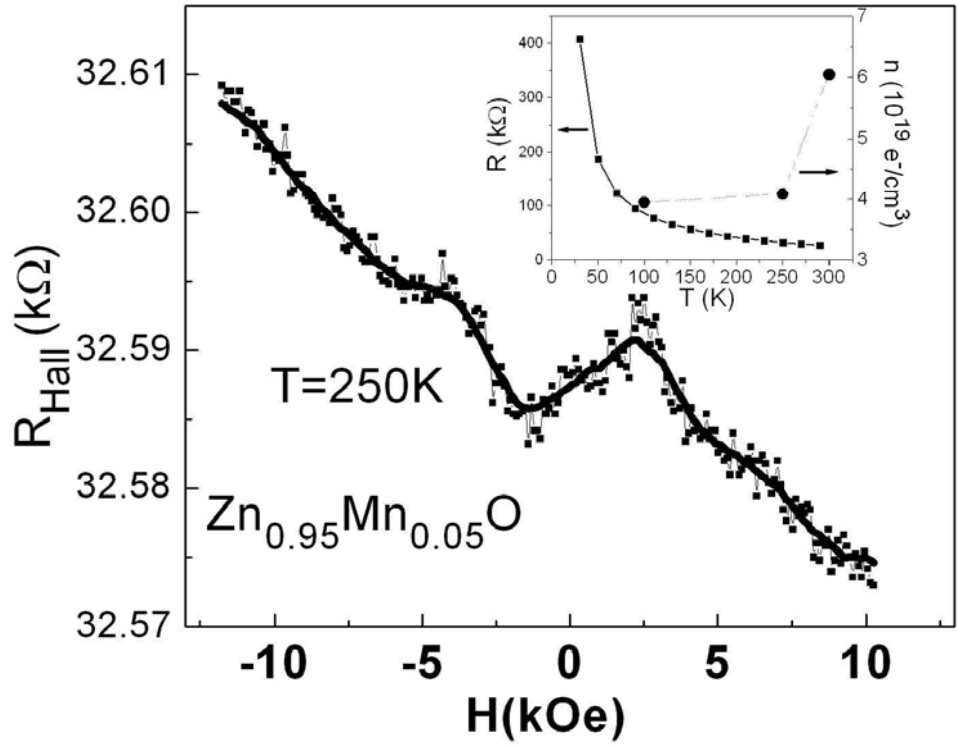

Figure 4